\documentclass[prb,twocolumn,superscriptaddress,showpacs,showkeys]{revtex4}

\usepackage{graphics}
\usepackage{color}
\usepackage{amsmath}
\usepackage{amssymb}
\usepackage[colorlinks=false]{hyperref}
\usepackage{pslatex}
\usepackage{amsfonts}
\usepackage{amsthm}
\usepackage{graphicx}

\include{MyMc} 

\begin{document}

\title{
  Oscillatory eigenmodes and stability of one and two arbitrary fractional vortices in long Josephson 0-$\kappa$-junctions
}

\author{E.~Goldobin}
\email{gold@uni-tuebingen.de}
\homepage{http://www.geocities.com/e_goldobin}
\affiliation{
  Physikalisches Institut II,
  Universit\"at T\"ubingen,
  Auf der Morgenstelle 14,
  D-72076 T\"ubingen, Germany
}

\author{H.~Susanto}
\email{h.susanto@math.utwente.nl}
\affiliation{Department of Applied Mathematics, University of
Twente, P.O. Box 217, 7500 AE Enschede, The Netherlands }

\author{D.~Koelle}
\author{R.~Kleiner}
\affiliation{
  Physikalisches Institut II,
  Universit\"at T\"ubingen,
  Auf der Morgenstelle 14,
  D-72076 T\"ubingen, Germany
}

\author{S. A.~van Gils}
\affiliation{Department of Applied Mathematics, University of
Twente, P.O. Box 217, 7500 AE Enschede, The Netherlands }




\date{\today}

\begin{abstract}
  We investigate theoretically the eigenmodes and the stability of one and two arbitrary fractional vortices pinned at one and two $\kappa$-phase discontinuities in a long Josephson junction. In the particular case of a single $\kappa$-discontinuity, a vortex is spontaneously created and pinned at the boundary between the 0 and $\kappa$-regions. In this work we show that only two of four possible vortices are stable. A single vortex has an oscillatory eigenmode with a frequency within the plasma gap. We calculate this eigenfrequency as a function of the fractional flux carried by a vortex. 
  
  For the case of two vortices, pinned at two $\kappa$-discontinuities situated at some distance $a$ from each other, splitting of the eigenfrequencies occur. We calculate this splitting numerically as a function of $a$ for different possible ground states. We also discuss the presence of a critical distance below which two antiferromagnetically ordered vortices form a strongly coupled ``vortex molecule'' that behaves as a single object and has only one eigenmode.

\end{abstract}

{\small
  To be submitted to Phys.\ Rev.\ B Regular (8 pages)
}

\maketitle

\section{Introduction}
\label{Sec:Intro}

Vortices in long Josephson junctions (LJJs) usually carry a single magnetic flux quantum $\Phi_0$ and therefore are often called fluxons. The study of dynamics of fluxons has been attracting a lot of attention during the last 40 years because of their interesting nonlinear nature\cite{Barone:JosephsonEffect,Likharev:JJ&Circuits,Ustinov:SolitonsLJJ} as well as because of potential applications\cite{VPK:IntRecv,Kemp:2002:VortexQubit}.

Recently it was shown that one can fabricate so-called 0-$\pi$-LJJs, \ie, LJJs consisting of alternating regions with positive (0-part) and negative ($\pi$-part) critical currents. Such junctions can be fabricated using superconductors with anisotropic order parameter which changes sign depending on the direction in $k$-space (\eg $d$-wave order parameter symmetry)\cite{Tsuei:Review,Smilde:ZigzagPRL} or with an oscillating order parameter (\eg with a ferromagnetic barrier)\cite{Bulaevskii:pi-loop,Ryazanov:2001:SFS-PiJJ,Kontos:2002:SIFS-PiJJ}. It was shown that at the boundary between a 0 and a $\pi$ part a new type of vortex carrying only half of the flux quantum may exist\cite{Bulaevskii:0-pi-LJJ,Xu:SF-shape,Kuklov:1995:Current0piLJJ}. Such vortices (naturally called semifluxons\cite{Goldobin:SF-Shape}) were observed experimentally in several types of 0-$\pi$-LJJs\cite{Kirtley:SF:HTSGB,Kirtley:SF:T-dep,Sugimoto:TriCrystal:SF,Hilgenkamp:zigzag:SF}. They may appear spontaneously and correspond to the ground state of the system\cite{Kogan:3CrystalVortices,Goldobin:SF-ReArrange,Zenchuk:2003:AnalXover}. 

The dynamics of the Josephson phase in a LJJ is described by a perturbed sine-Gordon equation. For the case of the 0-$\pi$-LJJ this equation is slightly modified\cite{Goldobin:SF-Shape} and includes the function $\theta(x)$ which is equal to zero all along the 0-parts and is equal to $\pi$ all along the $\pi$-parts. The function $\theta(x)$ as well as the Josephson phase $\phi(x,t)$, which is a solution of the sine-Gordon equation, is a $\pi$ discontinuous function of $x$ at the 0-$\pi$-boundaries. We sometimes call the 0-$\pi$-boundaries \emph{discontinuity points}. 

Recently, a LJJ geometry which allows to create \emph{arbitrary} discontinuities was suggested and successfully tested\cite{Goldobin:Art-0-pi}. In this LJJ a pair of closely situated current injectors creates an arbitrary $\kappa$-discontinuity (not only $\kappa=\pm\pi$) of the Josephson phase, with $\kappa$ being proportional to the current passing through the injectors\cite{Goldobin:Art-0-pi,Malomed:2004:ALJJ:Ic(Iinj)}. Since the Josephson phase is defined modulo $2\pi$ without loosing generality below we consider only $0\leq\kappa\leq 2\pi$.

Similar to the case of a $\pi$-discontinuity, the presence of an arbitrary $\kappa$-discontinuity results in formation of a fractional vortex pinned at it\footnote{%
Here we consider fractional vortices which appear as a result of the phase discontinuity\cite{Goldobin:SF-Shape,Goldobin:2KappaGroundStates}. Other types of arbitrary fractional vortices may appear in a LJJ with a strong second harmonic in the current-phase relation, which can be present either intrinsically or due to a faceted grain-boundary (frequently alternating regions of size $\ll\lambda_J$ of negative and positive critical current). In contrast to such ``splintered'' vortices\cite{Mints:2002:SplinteredVortices@GB,Mints:2002:NonLocal+FracVortices,Buzdin:2003:phi-LJJ} which are not pinned, the fractional vortices discussed here are pinned at a $\kappa$-discontinuity. The mixture of both possibilities, \ie, a fractional vortex at the boundary between a $\varphi$-region with the second harmonic and a 0-region without the second harmonic is also possible\cite{Buzdin:2003:phi-LJJ}%
}.
In the following, the \emph{topological charge} of a single vortex is defined to be a conserved quantity under the boundary condition imposed on the phase $\mu(x)$, and it is equal to the difference between the phase at $x=+\infty$ and $x=-\infty$. We use the word $\kappa$-vortex, to denote a vortex with the topological charge equal to $\kappa$. Given a discontinuity $-\kappa$, the topological charge of the vortex should be such that its sum with the discontinuity is equal to $2\pi n$ ($n$ is an integer), otherwise the energy of the system diverges\cite{Goldobin:2KappaGroundStates}. In fact one can construct not an infinite number but only four such vortices  corresponding to $n=-2,\,-1,\,0,\,1$, namely a $(\kappa-4\pi)$-vortex, a $(\kappa-2\pi)$-vortex, a $\kappa$-vortex, and a $(\kappa+2\pi)$-vortex\cite{Goldobin:2KappaGroundStates}.

It is shown below that only two of these vortices are stable: a so-called \emph{direct} $+\kappa$-vortex and a \emph{complementary} $(\kappa-\sgn(\kappa)2\pi)$-vortex. By definition the complementary of a complementary vortex, gives again a \emph{direct} $+\kappa$-vortex. The only exception is the case $\kappa=2\pi n$, for which there exist three stable solutions: a $+2\pi$-vortex (fluxon), the flat phase state (zero phase), and a $-2\pi$-vortex (antifluxon). The complementary ``vortex'' for both $\pm2\pi$-vortices is a constant phase state and, vice versa, for the constant phase state the complimentary vortex is either a fluxon or an antifluxon. In the majority of situations this can be distinguished due to the conservation of the topological charge.

Since, for $\kappa\ne 2\pi n$, $\kappa$-vortices are pinned, they might have an \emph{eigenmode} corresponding to the oscillation of the flux around the discontinuity point. Imagine that we apply a small uniform bias current through a LJJ containing a fractional vortex. The current creates a driving (Lorentz) force which bends the vortex. The driving force is compensated by the elastic (return) force which arises due to vortex deformation. If we now suddenly remove the bias current, the elastic force makes the vortex oscillate around the equilibrium shape with the eigenfrequency $\omega_0$, provided the damping is low enough (like in artificial 0-$\kappa$-LJJs\cite{Goldobin:Art-0-pi}).

We note that an integer fluxon \emph{does not have} any non-trivial eigenfrequencies\footnote{%
A fluxon has two trivial eigenvalues, \ie, at zero frequency (with multiplicity two) and at the edges of the plasma-band.\cite{Derks:sols} The zero eigenvalue is due to the translational invariance. The eigenvalue at the edges of the plasma-band bifurcates and becomes an internal mode in the presence of a perturbative term in the sine-Gordon equation.%
}.
Once deformed or perturbed, a fluxon recovers its shape exponentially without any internal oscillations. When the bias current is applied, the fluxon starts moving in a certain direction (which depends on the polarity of the fluxon and the bias current) which formally corresponds to $\omega_0=0$ (dc motion).

In this paper we investigate the stability and the eigenmodes of the vortex states formed in a LJJ with one or two $\kappa$-discontinuities.
The paper is organized as follows. In section \ref{Sec:AnCalc} we present an analytical derivation of the eigenfrequency of a single $\kappa$-vortex and compare the obtained results with the results of direct numerical simulations of the sine-Gordon equation. Then we study numerically two coupled $\kappa$-vortices in section \ref{Sec:TwoVortices}. The eigenfrequency in some limiting cases is derived analytically . Section \ref{Sec:Conclusion} concludes this work.

\section{Eigenfrequency of a single vortex}
\label{Sec:AnCalc}

For our study, it is more convenient to use the continuous Josephson phase $\mu(x,t)$ defined as $\mu(x,t)=\phi(x,t)-\theta(x)$.\cite{Goldobin:SF-Shape} A perturbed sine-Gordon equation which describes the dynamics of the Josephson phase in a 0-$\kappa$-LJJ can be written in terms of $\mu(x,t)$ as\cite{Goldobin:SF-Shape}
\begin{equation}
  \mu_{xx}-\mu_{tt}-\sin[\mu+\theta(x)] = \alpha\mu_t-\gamma(x),
  \label{Eq:sG-mu-kappa}
\end{equation}
with
\begin{equation}
 \theta(x)=\left\{
 \begin{array}{ll}
   0,~~x<0,\\
   -\kappa,~~x>0
 \end{array}
 \right.
\label{Eq:theta}
\end{equation}
for the case of a single discontinuity at $x=0$. Without loss of generality, we assume $0\leq\kappa\leq 2\pi$. The Josephson phase $\mu(x,t)$ is a continuous function of the coordinate $x$ and time $t$, which are normalized to the Josephson penetration depth $\lambda_J$ and to the inverse plasma frequency $\omega_p^{-1}$, respectively; $\alpha=1/\sqrt{\beta_c}$ is the dimensionless damping parameter, $\beta_c$ is the McCumber-Stewart parameter, and $\gamma(x)=j(x)/j_c$ is the bias current density $j(x)$ normalized to the critical current density $j_c$. The subscripts $x$ and $t$ denote corresponding partial derivatives with respect to $x$ and $t$, accordingly.

It is natural to have the boundary conditions at $x=0$ given by
\begin{equation}
  \begin{array}{ll}
    \lim_{x\uparrow 0}\mu(x)= \lim_{x\downarrow 0}\mu(x),\\
    \lim_{x\uparrow 0}\mu_x(x)= \lim_{x\downarrow 0}\mu_x(x).
  \end{array}
  \label{Eq:bcs}
\end{equation}

The zero bias static solution of Eq.~(\ref{Eq:sG-mu-kappa}) corresponding to a direct $\kappa$-vortex is given by
\begin{equation}
\mu^\kappa(x) = \left\{
 \begin{array}{ll}
 \mu^\kappa_-(x)=f(x), \, x<0,\\
 \mu^\kappa_+(x)=\kappa-f(-x), \, x>0,
 \end{array}\right.
\label{Eq:frac-vort}
\end{equation}
with
\[
f(x)=4\tan^{-1}e^{(x+x_0)}.
\]
The constant $x_0$ still has to be determined.

Imposing the boundary conditions (\ref{Eq:bcs}) to
$\mu^\kappa(x)$, we end up with the expression for $x_0$:
\begin{equation}
  x_0=\ln\tan\frac{\kappa}{8}
  . \label{Eq:x0}
\end{equation}

Next we will calculate the eigenfrequency of $\mu^\kappa$. The calculation we present in this article follows the one from Ref.~\onlinecite{Derks04}.
First we linearize Eq.~(\ref{Eq:sG-mu-kappa}) about the solution $\mu^\kappa$. We write
$\mu(x,t)=\mu^\kappa+u(x,t)$ and substitute the spectral ansatz
$u=e^{\lambda t}v(x)$ into Eq.~(\ref{Eq:sG-mu-kappa}). Retaining the
terms linear in $u$ gives the following eigenvalue problem
\begin{equation}
  v_{xx}-\left[\lambda^2+\alpha\lambda+\cos(\mu^\kappa+\theta)\right]v=0
  .\label{Eq:EP}
\end{equation}
Note that $\gamma$ is assumed to be zero here.

If we define
\[
  v=\left\{ \begin{array}{ll}
  v^-,\,x<0,\\v^+,\,x>0\end{array}\right.
\]
then at $x=0$ the boundary conditions are
\begin{subequations}
  \begin{eqnarray}
    v^-(0) &=& v^+(0),
    \label{Eq:bcs2a}\\
    \partial_xv^-(0) &=& \partial_xv^+(0).
    \label{Eq:bcs2b}
  \end{eqnarray}
  \label{Eq:bcs2}
\end{subequations}

From Eq.~(\ref{Eq:EP}), we know that $v^\pm$ satisfies the equation
\[ 
  v^\pm_{xx}-
  \left(\lambda^2+\alpha\lambda+\cos(f(\mp x))\right)v^\pm=0.
\]
The eigenfrequency corresponds to a solution $v^\pm$ that tends to $0$ as $x\to\pm\infty$. According to Refs.~\onlinecite{Derks:sols} and \onlinecite{Mann:sols} the above eigenvalue problem, if considered on the whole real line, has two independent solutions. We will use these known solutions to construct the eigenfunction and find the corresponding eigenvalue for our problem, \ie, including the discontinuity. The solution that corresponds to the eigenfunction is given by
\begin{equation}
  v = \left\{
    \begin{array}{ll} 
      v^-(x)=e^{\Lambda (x+x_0)}(\tanh(x+x_0)-\Lambda);\\
      v^+(x)=v^-(-x),
    \end{array}
  \right.
  \label{ln0}
\end{equation}
where 
\begin{equation}
  \Lambda=\sqrt{\lambda^2+\alpha\lambda+1}
  . \label{Eq:Lambda}
\end{equation}
The condition that $v^\pm\to 0$ as $x\to \pm\infty$ requires
that the real part of lambda, $\Re(\Lambda)$, is positive. The eigenvalue $\lambda$ is calculated from $\Lambda$ by determining that $v$ has to
satisfy Eqs.~(\ref{Eq:bcs2}).

From Eq.~(\ref{Eq:bcs2b}) we obtain that
\[
  \begin{array}{ll}
    v^+_x(0)-v^-_x(0)=2\left.e^{\Lambda y}(\Lambda(\tanh
    y-\Lambda)+\sech^2y)\right|_{x=0}=0,\\y=\pm x+x_0,
  \end{array}
\]
from which we obtain
\[
  \Lambda=\frac12\left(\tanh x_0\pm\sqrt{\tanh^2x_0+4\sech^2x_0}\right).
\]
When $\alpha=0$, the eigenfrequency $\omega_0$ is given by
\begin{equation}
  \lambda=i\omega_0=\pm\frac{i}{2}\sqrt{4-\left(\tanh
  x_0+\sqrt{\tanh^2 x_0+4\sech^2x_0}\right)^2}
  ,\label{Eq:EF}
\end{equation}
with $x_0$ given by Eq.~(\ref{Eq:x0}). This can be further simplified to give
\begin{equation}
  \omega_0(\kappa)=\sqrt{\frac12\cos\frac\kappa4
  \left( \cos\frac\kappa4 + \sqrt{4-3\cos^2\frac\kappa4} \right)
  }
  . \label{Eq:EigenFreq}
\end{equation}

\begin{figure}[bt]
  \begin{center}
    \includegraphics{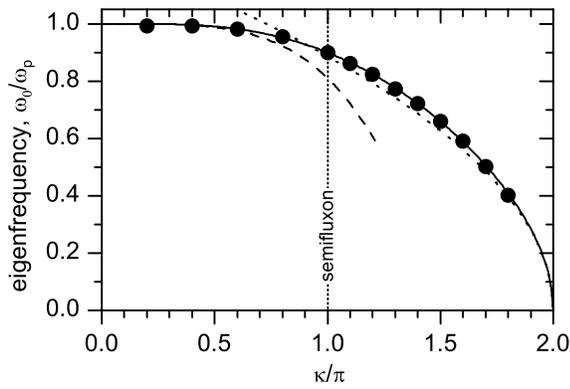}    
  \end{center}
  \caption{%
    Plot of the normalized eigenfrequency $\omega_0$ of a single $\kappa$-vortex versus $\kappa$ given by Eq.~(\ref{Eq:EigenFreq}) [continuous line] and its approximations for $\kappa\to0$: $\omega_0(\kappa)\approx1-\kappa^4/512$ [dashed line] and for $\kappa\to2\pi$: $\omega_0(\kappa)\approx\frac12\sqrt{2\pi-\kappa}$ [dotted line]. Symbols show the values of eigenfrequency obtained by direct numerical simulation of Eq.~(\ref{Eq:sG-mu-kappa}).}
  \label{Fig:EigenFreq}
\end{figure}

The plot of this eigenfrequency as a function of $\kappa$ is shown in Fig.~\ref{Fig:EigenFreq}. For the particular case of a semifluxon the eigenfrequency is $\omega_0(\pi)=\frac12\sqrt{1+\sqrt5}\approx0.899$. Note that $\omega_0(\kappa)=\omega_0(-\kappa)$ because the $\kappa$-vortex and the $-\kappa$-vortex have the same eigenfrequencies.

The above result is the only eigenvalue of a direct fractional vortex. For a rigorous proof, we refer to Ref.~\onlinecite{Derks:sols}. Below we will show that there is no eigenfunction with at least one zero at finite $x$. 

According to the Sturm-Liouville theorem, if Eq.~(\ref{Eq:EP}) has several eigenvalues $E=\lambda^2+\alpha\lambda=\Lambda^2-1$ given by, \eg, $E_1,\,E_2,\,\dots,\,E_n$, then we can arrange them such that $E_1<E_2<\dots<E_n$. The eigenfunction corresponding to $E_n$ will have $(n-1)$ zeros.

Next, based on the first line of Eq.~(\ref{ln0}), we construct an anti-symmetric solution with $v^+(x)=-v^-(-x)$. 
With this property of $v$, $v=0$  at $x=0$ and Eq.~(\ref{Eq:bcs2}) can be fulfilled if and only if 
\begin{equation}
  v(0)=0
  \quad\text{or}\quad
  \Lambda=\tanh\left(x_0\right) 
  = -\cos\left(\frac{\kappa}{4}\right) \leq 0.
  \label{v0}
\end{equation}
We see that there is no solution $\Lambda$ with positive real part, which is necessary to let $v^\pm\to 0$ as $x\to \pm\infty$. Only at $\kappa=2\pi$ we have an eigenvalue attached to the edge of the continuous spectrum and the corresponding eigenfunction is bounded, but not exponentially decaying.\cite{Derks:sols} At this value of $\kappa$, the solitary wave we consider is nothing else but an integer fluxon. Hence, a stable fluxon has no internal mode.

We have also checked our analytical results numerically using {\sc StkJJ} software\cite{StkJJ}. We simulated a rather long $L=20\lambda_J$ Josephson junction with a $\kappa$-discontinuity at its center $x=0$. 
To obtain the eigenfrequency, we applied a small bias current which pulls the $\kappa$-vortex in a certain direction and waited to arrive to a stable stationary state. Then we abruptly decreased the bias current back to zero and observed the $\mu_t(t)\propto V(t)$ at $x=1$, where $V(t)$ is the normalized voltage. The plot $V(t)$ exhibits decaying oscillations. Making a Fourier transform, we find a single dominant frequency which is considered as an eigenfrequency of the $\kappa$-vortex. We repeated this procedure for different positions $x$ and different small initial bias currents, but the results did not change within 1\% of accuracy. Repeating this numerical experiment for several $\kappa$ in the range from $0$ to $2\pi$ we obtained the values of eigenfrequencies which are shown as filled symbols in Fig.~\ref{Fig:EigenFreq}, demonstrating perfect agreement between the analytics and the direct simulations. The eigenfrequency of the complimentary $(\kappa-2\pi)$-vortex can be obtained by using $2\pi-\kappa$ instead of $\kappa$ in Eq.~(\ref{Eq:EigenFreq}). 

Note that Eq.~(\ref{Eq:EF}) is also valid for $(\kappa+2\pi n)$-vortices with the topological charge larger than $2\pi$. However, when $|\kappa+2\pi n|>4\pi$ the static solution cannot be constructed\cite{Goldobin:2KappaGroundStates}. Therefore we consider here only the cases $n=1$ and $n=-2$. The zero bias static solution of Eq.~(\ref{Eq:sG-mu-kappa}) corresponding to a fractional $(2\pi n+\kappa)$-vortex is given by
\begin{equation}
  \mu^\kappa(x) = \left\{
    \begin{array}{ll}
      \mu^\kappa_-(x)=f(x), \, x<0,\\
      \mu^\kappa_+(x)=2\pi n+\kappa-f(-x), \, x>0,
    \end{array}\right.
  \label{Eq:2pi+kappa}
\end{equation}
with
\[
  f(x)=4\tan^{-1}e^{(x+x_n)}
  ,
\]
where $x_n$ is determined by matching the boundary conditions (\ref{Eq:bcs2}) as described above. Thus,
\begin{equation}
  x_n=\ln\tan\left(\frac{\pi n}{4}+\frac{\kappa}{8}\right)
  . \label{Eq:x1}
\end{equation}
The eigenfrequency of this state is still given by (\ref{Eq:EF}) with $x_0$ being substituted by $x_n$. One can check by direct substitution of (\ref{Eq:x1}) into (\ref{Eq:EF}) that $\Re(\lambda)>0$ for $n=+1$ or $n=-2$, \ie, the $(\kappa+2\pi)$ and $(\kappa-4\pi)$ vortices are unstable and, therefore, not observable. A damping cannot stabilize these ``heavy'' vortices.

If one thinks about the dispersion relation $\omega(k)=\sqrt{1+k^2}$ of a LJJ, the dispersion relation has a gap (plasma gap) from $\omega=0$ to $\omega=1$. In the presence of a vortex, there is an additional discrete level (eigenfrequency) situated within the plasma gap and corresponding to $k=0$ (localized state). This level is somewhat similar to impurity levels in semiconductors. The position of this ``impurity'' level can be tuned electronically by changing $\kappa$.

\section{Two coupled vortices}
\label{Sec:TwoVortices}

Now consider two $\kappa$-discontinuities ($0<\kappa<2\pi$) in an infinitely long LJJ at the distance $a$ from each other (at $x=\pm a/2$). If both discontinuities have the same sign of $\kappa$ (\eg $+\kappa$), there are two possible irreducible vortex configurations: the symmetric ferromagnetic (FM) state $\stat{dd}=(-\kappa,-\kappa)$, and the asymmetric antiferromagnetic (AFM) state $\stat{dU}=(-\kappa,2\pi-\kappa)$\cite{Goldobin:2KappaGroundStates}. If the discontinuities have different sign, \eg, $+\kappa$ and $-\kappa$, there are two other irreducible vortex states: the asymmetric FM state $\stat{Uu}=(2\pi-\kappa,\kappa)$, and the symmetric AFM state $\stat{du}=(-\kappa,+\kappa)$\cite{Goldobin:2KappaGroundStates}. The details about ground states are presented in Ref.~\onlinecite{Goldobin:2KappaGroundStates}. Below we consider eigenmodes of these ground states. 

Note that in the notations such as \state{udUD} the direction of the arrows shows the polarity of the vortex (up or down), while the harpoon on the left or on the right side indicates whether the vortex is direct or complementary. The notations are summarized in Tab.~\ref{Tab:Notations}.

\begin{table}
  \begin{tabular}{cccl}
    \textbf{Symbol} & \textbf{Discontinuity} & 
    \textbf{Topological charge} & \textbf{Name}\\
    \hline
    \state{u} & $-\kappa$ & $+\kappa$ & direct\\
    \state{d} & $+\kappa$ & $-\kappa$ & direct\\
    \state{U} & $+\kappa$ & $2\pi-\kappa$ & complementary\\
    \state{D} & $-\kappa$ & $\kappa-2\pi$ & complementary\\
    \state{s} & $\pm\pi$ & $+\pi$ & semifluxon\\
    \state{a} & $\pm\pi$ & $-\pi$ & antisemifluxon\\
    \state{F} & $0$ & $+2\pi$ & fluxon\\
    \state{A} & $0$ & $-2\pi$ & antifluxon\\
    \hline
  \end{tabular}
  \caption{%
    Notations for different types of vortices.
  }
  \label{Tab:Notations}
\end{table}

In a system of two weakly interacting vortices the eigenfrequency $\omega_0$ splits into two different frequencies: $\omega_+$, corresponding to the in-phase oscillations of both fractional vortices, and $\omega_-$, corresponding to the anti-phase oscillations. By definition adopted here the ``in-phase'' or ``anti-phase'' means that \emph{magnetic fields} (rather than Josephson phases) of two vortices oscillate in-phase or anti-phase. 
According to the Sturm-Liouville theorem the eigenfunctions $\nu(x)$ (oscillations of the Josephson phase) corresponding to the first and the second lowest eigenfrequencies have no zeros and one zero, respectively. In turn, the oscillations of magnetic field $\nu_x(x)$ have 1 or 2 zeros. If one draws such $\nu_x(x)$ deviations qualitatively on the top of the magnetic field profiles $\mu_x(x)$, corresponding to various ground states, one can conclude which eigenfrequency corresponds to in-phase and which to anti-phase oscillations.

In general, it is quite difficult to calculate the splitting of the eigenfrequency analytically\footnote{The splitting of an eigenvalue out of the edges of the phonon band cannot be calculated using the formal expansion method. This is an indication of an exponentially small splitting of the eigenvalue.}, so below we calculate the frequencies $\omega_\pm(\kappa)$ for different values of $a$ numerically. 

To obtain the data presented below we have used a technique similar to the one mentioned in Sec.~\ref{Sec:AnCalc}. To excite in-phase oscillations we applied a small uniform bias current. To excite anti-phase oscillations the bias current $\gamma(x)$ was applied so that $\gamma(x)>0$ for $x>0$ and $\gamma(x)<0$ for $x<0$. We used several different $\gamma(x)$, \eg, $\gamma(x)\propto x$ or $\gamma(x)\propto\sgn(x)$, and the results were generally the same. 

Unfortunately this technique is very slow, requires a lot of manual work and is not very precise. Therefore, we implemented another technique for calculating the eigenvalues in StkJJ\cite{StkJJ}. When the numerical solution decays to a stable stationary one, we analyze its stability, by introducing an arbitrary perturbation $\delta$ to the found solution $\mu_0$, \ie, we substitute $\mu=\mu_0+\delta$ into (\ref{Eq:sG-mu-kappa}) and obtain a PDE for $\delta$. Then we discretize the obtained PDE and arrive to a system of $N$ coupled second order ODEs. Typically for a LJJ of length $L=40\lambda_J$, we use the spatial discretization step $\Dx=0.02\lambda_J$ which gives us $N=L/\Dx=2000$. This system can, in turn, be reduced to a system of $2N$ ODEs of the first order, which can be written in a matrix form as
\begin{equation}
  \dot{\mathbf{\delta}}=A\mathbf{\delta},
  \label{equation}
\end{equation}
where $\delta$ is a vector with $2N$ elements and $A$ is a constant matrix constructed using $\mu_0$. The eigenvalues of this matrix can be found using standard routines\cite{NRC++}. Among all $2N$ complex eigenvalues we chose two eigenvalues with the smallest imaginary part (eigenfrequency). The other eigenvalues had $|\Im(\lambda)|>1$, \ie, belonged to the plasma band. The real parts of all eigenvalues were negative (and $\approx-\alpha/2$), indicating the stability of the solution. In this way, we were able to calculate eigenfrequencies numerically, with high accuracy for hundreds of $\kappa$ values in automatic fashion. The results produced by both methods were compared in a number of selected points and were essentially the same.

\subsection{Symmetric AFM state}
\label{Sec:sAFM}

\begin{figure}[!tb]
  \centering
  \includegraphics{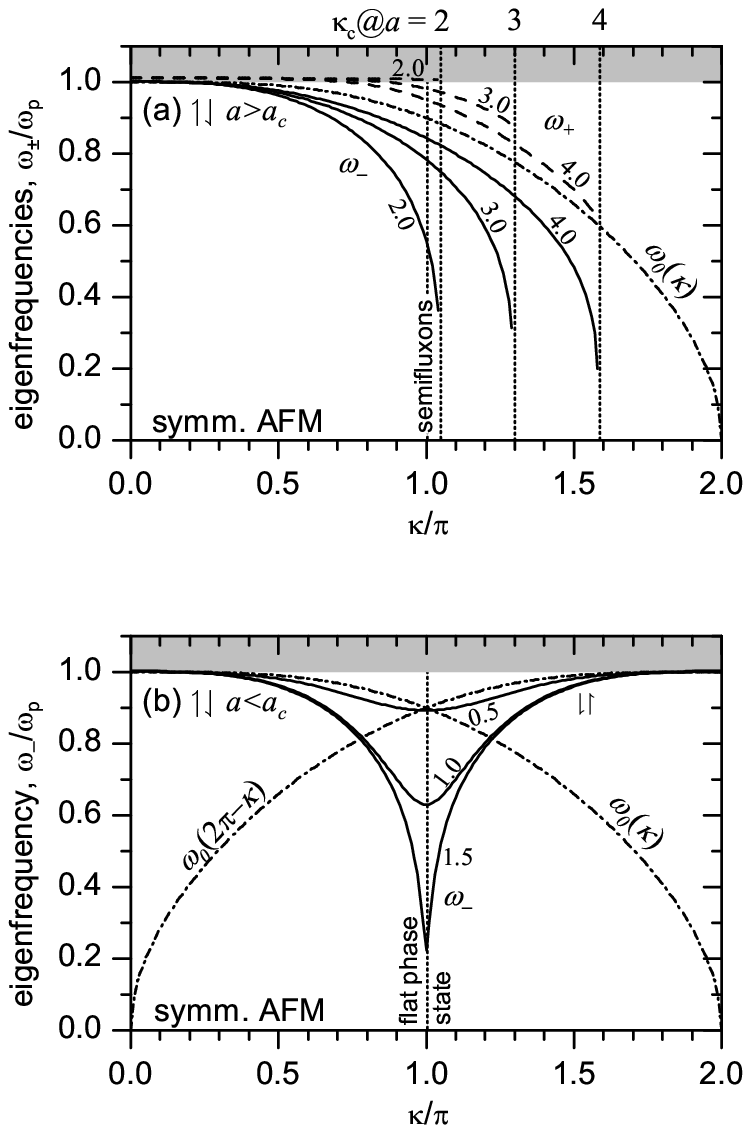}
  \caption{%
    Eigenfrequencies $\omega_{-}(\kappa)$ (solid line) and $\omega_{+}(\kappa)$ (dashed line) for the symmetric AFM state calculated numerically for different distances $a$ between the vortices in the limit of (a) weak ($a>a_c$) and (b) strong ($a<a_c$) coupling. For the case $a>a_c$ only the state \protect\state{ud} is shown. For $a<a_c$ both states \protect\state{ud} ($\kappa<\pi$) and \protect\state{DU} ($\kappa>\pi$) are shown. Dashed-dotted lines show eigenfrequencies for a single direct or complimentary vortex in a LJJ with one discontinuity.
  }
  \label{Fig:SymmAFM:omega_pm}
\end{figure}

First, let us consider a LJJ with two $(+\kappa, -\kappa)$ discontinuities and with $(-\kappa, +\kappa)$ vortices \state{du} pinned at them. Since this symmetric AFM state is the most natural state of the system, in simulations we were simply sweeping $\kappa$ from 0 to $2\pi$ starting from $\mu=0$ solution at $\kappa=0$ and the symmetric AFM state was formed automatically. 
The numerically obtained eigenfrequencies $\omega_\pm(\kappa)$ for different distances $a$ between vortices are shown in Fig.~\ref{Fig:SymmAFM:omega_pm}. 

It was shown\cite{Goldobin:2KappaGroundStates} that there is a critical value of discontinuity $\kappa_c^{\state{ud}}(a)$ ($\pi<\kappa_c^{\state{ud}}<2\pi$) at which the ground state \state{ud} switches to the state \state{DU} ($\kappa-2\pi$ and $2\pi-\kappa$ vortices pinned at $-\kappa$ and $+\kappa$ discontinuities). 
The inverse function for $\kappa_c^{\state{ud}}(a)$ is denoted as $a_c^{\state{ud}}(\kappa)$.\cite{Goldobin:2KappaGroundStates}

When $a>a_c^{\state{ud}}(\pi)=a_c=\pi/2$, see Fig.~\ref{Fig:SymmAFM:omega_pm}(a), the splitting of the eigenfrequency is clearly visible with $\omega_-<\omega_0<\omega_+$. The eigenfrequencies $\omega_\pm$ are calculated for $0<\kappa<\kappa_c^{\state{ud}}(a)$. At the critical value $\kappa_c^{\state{ud}}(a)$, the state \state{ud} becomes unstable. As a sign of this instability one can see that $\omega_-$ sharply approaches 0 as $\kappa\to\kappa_c$. 

The eigenfrequency of the complementary state $\omega^{\state{UD}}_\pm(\kappa)$ is given by
\[
  \omega^{\state{UD}}_\pm(\kappa)=\omega^{\state{ud}}_\pm(2\pi-\kappa)
  ,\quad \kappa>2\pi-\kappa_c.
\]
Note, that in the interval of $\kappa$ from $2\pi-\kappa_c^{\state{ud}}$ to $\kappa_c^{\state{ud}}$ both \state{ud} and \state{UD} states are stable, but have different energy\cite{Goldobin:2KappaGroundStates}. Hence, we have a bistability phenomenon. 

When $a<a_c$, see Fig.~\ref{Fig:SymmAFM:omega_pm}(b), the transition from the state \state{ud} to the state \state{DU} is smooth without hysteresis and happens at $\kappa=\pi$ via the flat phase state.\cite{Goldobin:2KappaGroundStates} Note that for $a<a_c$ the frequency $\omega_+=1$, \ie, belongs to the plasma band. This means that strongly coupled vortices behave like a single object with a single eigenfrequency $\omega_-$. When $a\to a_c$, $\omega_-(\pi)\to0$, corresponding to the instability of the flat phase solution in favor of the AFM state.\cite{Kato:1997:QuTunnel0pi0JJ,Goldobin:SF-ReArrange,Zenchuk:2003:AnalXover} For $a\to 0$, $\omega_-(\pi)\to1$ that corresponds to the absence of the eigenmodes when discontinuities cancel each other and one ends up with a conventional LJJ. In particular, for $\kappa=\pi$ and $a<a_c$ (flat phase state) one can calculate analytically the eigenvalue as a function of the facet length $a$. Note that the linearized equation describing the eigenvalue problem is nothing else but the Schr\"odinger equation with a potential well\cite{Kato:1997:QuTunnel0pi0JJ}. Some simple algebraic calculations give
\begin{equation}
  \lambda^2~\sin\left(a\sqrt{1-\lambda^2}\right)
  +\sqrt{1-\lambda^4}~\cos\left(a\sqrt{1-\lambda^2}\right) = 0
  . \label{linear_ac}
\end{equation}
Writing $\lambda=i\omega_-$, we will obtain $\omega_-(\pi)$ as the first root of Eq.~(\ref{linear_ac}). One can show as well that for $0<a<a_c$, $\omega_-(\pi)$ is the only eigenvalue of the system in accordance with the reported numerical results above.

\subsection{Symmetric FM state}
\label{Sec:sFM}

Now we consider a LJJ with two $(+\kappa, +\kappa)$ discontinuities with $(2\pi-\kappa, 2\pi-\kappa)$ vortices pinned at them, \ie, the \state{UU} state. The same applies to the \state{dd} state by substitution $\kappa\to2\pi-\kappa$. In simulations we were starting from the two semifluxon state $\state{ss}\equiv\state{UU}$ at $\kappa=\pi$ and were sweeping $\kappa$ towards 0 or towards $2\pi$. In a symmetric FM state the in-phase mode frequency $\omega_+$ is smaller than anti-phase mode frequency $\omega_-$, namely $\omega_+<\omega_0$ and $\omega_->\omega_0$. Below we calculate the frequencies $\omega_\pm$ for $\kappa_c(a)<\kappa<2\pi$. 

\begin{figure}[!tb]
  \centering
  \includegraphics{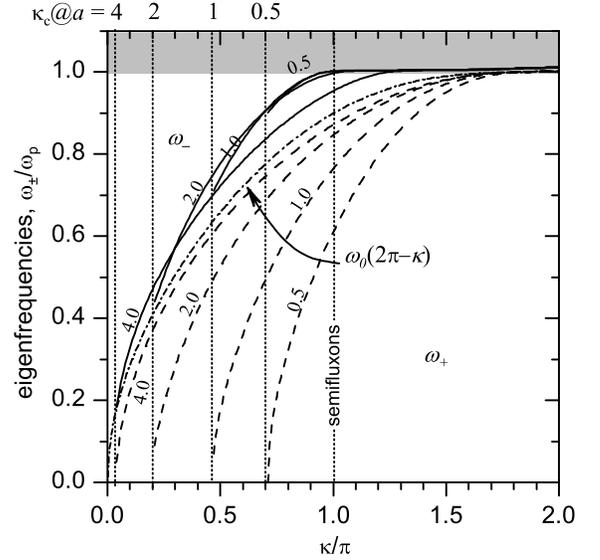}
  \caption{%
    Eigenfrequencies $\omega_{-}(\kappa)$ (solid lines) and $\omega_{+}(\kappa)$ (dashed lines) for the symmetric FM state \protect\state{UU} calculated numerically for different distances $a$ between the vortices. The dotted lines show the critical values of $\kappa_c$ for given $a$ at which the state \protect\state{UU} ceases to exist. Dashed-dotted lines show eigenfrequencies for a single direct or complimentary vortex in a LJJ with one discontinuity.
  }
  \label{Fig:SymmFM:omega_pm}
\end{figure}

The quantity $\kappa_c^{\state{UU}}(a)$ ($0<\kappa_c^{\state{UU}}<\pi$) is the critical value of discontinuity at which the ground state \state{UU} ceases to exist (becomes unstable) and the system switches to another state, \eg, \state{Ud}\cite{Goldobin:2KappaGroundStates}. The values of $\omega_\pm(\kappa)$ calculated numerically are shown in Fig.~\ref{Fig:SymmFM:omega_pm}. Note that the lower eigenfrequency $\omega_+(\kappa)\to0$ when $\kappa\to\kappa_c^{\state{UU}}$. 

In the limit of $a\to 0$, one can get an analytic expression of the eigenvalue as a function of $\kappa$. This is caused by the fact that in the limit $a\to0$, the two fractional vortices form a single vortex with a double topological charge. Hence, in this limit the largest eigenfrequency is given by $\omega_-=\omega_0(2\kappa)$ or, following the choice of vortices presented in Fig.~\ref{Fig:SymmFM:omega_pm}, $\omega_-=\omega_0(4\pi-2\kappa)$ with $\omega_0$ is given by Eq.~(\ref{Eq:EigenFreq}). In the case of $\kappa=\pi$, $\omega_-=0$ in agreement with the fact that the two semifluxons form an integer fluxon. 

Another property that is known in the limit of $a\to0$ is that there is no eigenvalue other than $\omega_-$ (see the calculations presented in Sec.~\ref{Sec:AnCalc}). The internal mode $\omega_+$ enters the phonon band in this limit leaving $\omega_+(\pi)=1$.

\subsection{Asymmetric AFM state}

\begin{figure}[!b]
  \centering\includegraphics{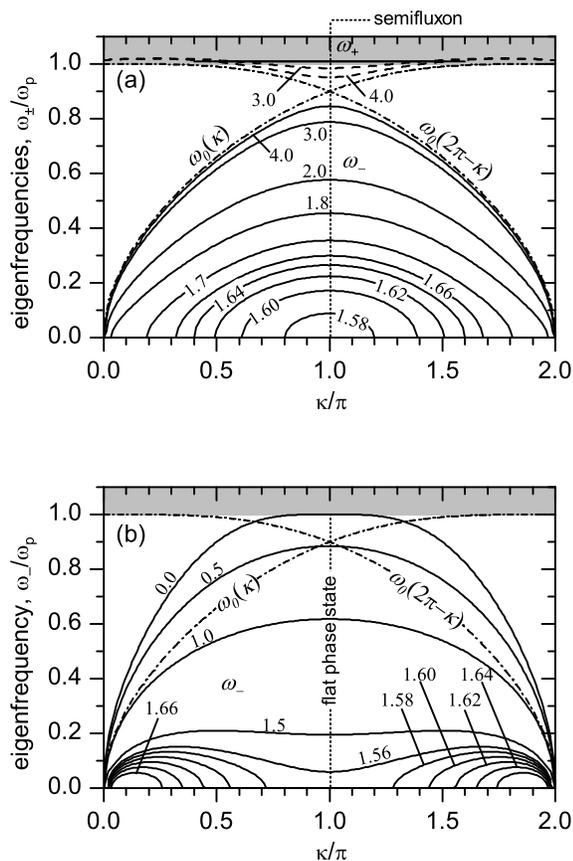}
  \caption{%
    Eigenfrequencies $\omega_{-}(\kappa)$ (solid line) and $\omega_{+}(\kappa)$ (dashed line) for the asymmetric AFM state \protect\state{dU} calculated numerically for different distances $a$ between vortices. In (a) and (b) the coupling between the two vortices is strong and weak, respectively. The line marked as 0.0 in (b) shows the behavior of the eigenfrequency in the limit $a\to0$ calculated analytically using Eq.~(\ref{Eq:AsymAFM@a=0}). Dashed-dotted lines show eigenfrequencies for a single direct or complimentary vortex in a LJJ with one discontinuity.
  }
  \label{Fig:AsymAFM}
\end{figure}

Now we consider two vortices with topological charges $-\kappa$ and $2\pi-\kappa>0$ in the AFM state \state{dU} admitted by $(+\kappa,+\kappa)$-discontinuities. In simulations we were starting from the state $\state{as}\equiv\state{dU}$ at $\kappa=\pi$ and were sweeping $\kappa$ towards 0 or towards $2\pi$. The calculated frequencies $\omega_\pm$ for $0<\kappa<2\pi$ are shown in Fig.~\ref{Fig:AsymAFM}. 

Note that although this state is asymmetric (at least for large $a$) in terms of magnetic flux carried by the vortices, its $\omega_\pm(\kappa)$ dependence is symmetric with respect to the $\kappa=\pi$ axis, \ie, $\omega_{\pm}^{\state{dU}}(\kappa)=\omega_{\pm}^{\state{Ud}}(\kappa)$. This state is stable for $0<\kappa<2\pi$. 

When the vortices are weakly coupled [$a>a_c^{\state{dU}}(\kappa)=a_c^{\state{uD}}(\kappa)$, Fig.~\ref{Fig:AsymAFM}(a)] we indeed have an asymmetric state\cite{Goldobin:2KappaGroundStates} and splitting of eigenfrequencies with $\omega_-<\omega_0<\omega_+$. In the opposite limit when the coupling is strong [$a<a_c^{\state{dU}}(\kappa)$, Fig.~\ref{Fig:AsymAFM}(b)], a new \emph{collective state} is formed.\cite{Goldobin:2KappaGroundStates} This state has only one eigenfrequency $\omega_{-}$, while $\omega_+=1$ and belongs to the plasma band. 

The crossover distance $a_c^{\state{dU}}(\kappa)$ which separates an asymmetric state from a collective state is a weak function of $\kappa$ as can be seen in Fig.~\ref{Fig:AsymAFM}. The crossover distance $a_c^{\state{dU}}(\kappa)$ grows from $a_c^{\state{dU}}(\pi)=\pi/2\approx1.57$ to $a_c^{\state{dU}}(0)=a_c^{\state{dU}}(2\pi)=2\ln(1+\sqrt2)\approx1.7628$.\cite{Goldobin:2KappaGroundStates} Thus, for $\pi/2<a<2\ln(1+\sqrt2)$ the asymmetric flux state exists not in the whole interval of $0<\kappa<2\pi$, but only in some smaller sub-interval which includes $\kappa=\pi$. In the same time the collective state nucleates around $\kappa=0$ and $\kappa=2\pi$. At $a \leq \pi/2$ the island of the asymmetric state vanishes and the collective state exists in the whole interval of $0<\kappa<2\pi$. Vice versa, at $a\geq 2\ln(1+\sqrt2)$ the asymmetric state exists for all $0<\kappa<2\pi$.

Since for $\kappa=\pi$ the symmetric and asymmetric AFM states are equivalent to the flat phase state, the same limiting behavior which was discussed at the end of Sec.~\ref{Sec:sAFM} apply [see also Eq.~(\ref{linear_ac}) and \cf Fig.~\ref{Fig:SymmAFM:omega_pm}].

It is interesting to follow the smallest eigenfrequency $\omega_-(\kappa)$ at $\kappa=\pi$ in Fig.~\ref{Fig:AsymAFM}. In Fig.~\ref{Fig:AsymAFM}(a), decreasing the facet length $a$ gives smaller $\omega_-(\pi)$ until the eigenfrequency reaches 0 at $a=a_c$ that corresponds to the transition to the flat phase state. After $a$ passes $a_c$, in the flat phase state, further decrease of the facet length increases $\omega_-(\pi)$. 

For arbitrary $\kappa$ and $a\to0$ the discontinuities $(+\kappa,+\kappa)$ become a single discontinuity $+2\kappa$. This case is similar to the limiting case we presented in Sec.~\ref{Sec:sFM}, but the topological charge of the fractional vortex is now $2\pi-2\kappa$. Hence, for $a\to0$ the eigenfrequency of the collective state approaches 
\begin{equation}
  \omega_-(\kappa)=\omega_0(2\pi-2\kappa)
  , \label{Eq:AsymAFM@a=0}
\end{equation}
with $\omega_0$ defined by Eq.~(\ref{Eq:EigenFreq}) [see the line marked as 0.0 in Fig.~\ref{Fig:AsymAFM}(b)].

\subsection{Asymmetric FM state}

\begin{figure}[!tb]
  \centering\includegraphics{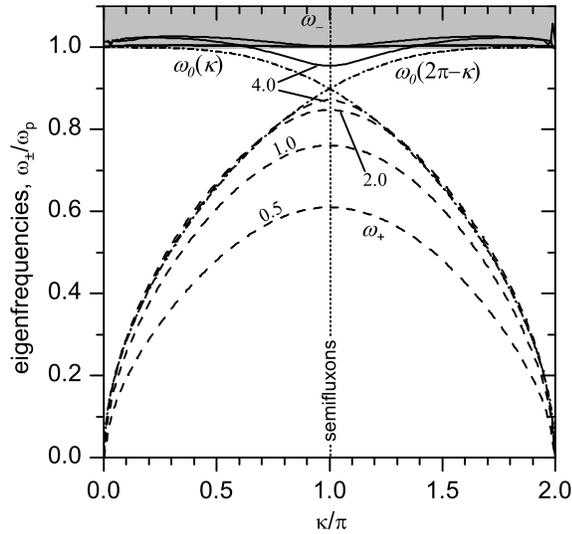}
  \caption{%
    Eigenfrequencies $\omega_{-}(\kappa)$ (solid line) and $\omega_{+}(\kappa)$ (dashed line) for the asymmetric FM state \protect\state{uU} calculated numerically for different distances $a$ between vortices. Dashed-dotted lines show eigenfrequencies for a single direct or complimentary vortex in a LJJ with one discontinuity.
  }
  \label{Fig:AsymFM}
\end{figure}

Finally, we consider two vortices $+\kappa$ and $2\pi-\kappa>0$ in the FM state \state{uU} admitted by two discontinuities $-\kappa$ and $+\kappa$. In simulations we were starting from the state $\state{ss}\equiv\state{uU}$ at $\kappa=\pi$ and were sweeping $\kappa$ towards 0 or towards $2\pi$. In this state
the in-phase mode frequency $\omega_+<\omega_0<\omega_-$. The calculated frequencies $\omega_\pm$ are shown in Fig.~\ref{Fig:AsymFM}. Again, the plot is symmetric with respect to $\kappa=\pi$, \ie, $\omega_\pm^{\state{Uu}}(\kappa)=\omega_\pm^{\state{uU}}(\kappa)$.

The asymmetric FM state has no crossovers and at $a\to0$ turns itself into a $2\pi$ vortex.

One can also recognize that for $a\ll 1$, ($-\pi,+\pi$)-discontinuities in a Josephson junction behave like a micro-resistor to an integer $2\pi$-fluxon (see, \eg, Eq.~(1.2) of Ref.~\onlinecite{Kato:1996:MQT-FluxonLJJ}). In the limit $a\to0$, the micro-resistor strength $\epsilon$ from Ref.~\onlinecite{Kato:1996:MQT-FluxonLJJ} is equal to $2a$.

\section{Conclusions}
\label{Sec:Conclusion}

We have shown that an arbitrary fractional vortex has an eigenmode that corresponds to the oscillations of the vortex around the phase discontinuity point where it is pinned. We have derived the eigenfrequency for a direct and a complimentary vortex as a function of the topological charge (magnetic flux) carried by the vortex. The eigenfrequency corresponds to a discrete ``impurity'' energy level within the plasma gap.

For the case of two coupled vortices we showed that the eigenfrequency splits into two modes that correspond to the in-phase and anti-phase oscillations of the vortices. Both of these eigenfrequencies were calculated numerically for each ground state. For $a<a_c$, two AFM ground states form a strongly coupled ``vortex molecule'' which behaves as a single object and has only one eigenmode.

The obtained results may have enormous impact on the study of vortex dynamics in novel Josephson systems, \eg, based on SFS junctions or $d$-wave superconductors. The knowledge of eigenmodes is a key element in designing classical or quantum devices based on fractional vortices. In the classical domain this may help to avoid parasitic resonance phenomena. In the quantum domain, the eigenfrequency determines the characteristic frequency of, \eg, quantum tunneling/flipping processes.

To detect eigenmodes experimentally one may use spectroscopic methods or resonant excitation. To make spectroscopy one should excite the eigenmode by applying an ac bias current or sending an electromagnetic wave along the LJJ from the edge. Then one should measure \eg the critical current as a function of the external excitation frequency $\omega$. The ac bias current can be applied by coupling the junction bias leads to a microwave antenna (and avoiding the induction of the microwave signal in all other parts, \eg, injectors) or by embedding the junction into a resonator. If we choose to send the microwaves from the edge, the length of the junction should not be very large and the frequency should not be very low, so that the condition 
\[
  \max\left[ \alpha,\sqrt{1-\left( \frac{\omega}{\omega_p} \right)^2} \right] \frac{L}{\lambda_J}<1
\] 
is satisfied, \ie, the microwave signal should not decay substantially while it propagates from the edge to the middle of the junction where the vortex (molecule) is created. Spectroscopic methods may help to distinguish between a direct $\kappa$-vortex and its complimentary $(2\pi-\kappa)$-vortex by measuring the eigenfrequency (except $\kappa\approx\pi$).

Another possibility to detect an eigenmode is to create a fractional vortex in an annular LJJ and let the integer fluxon, which can be injected if needed\cite{Ustinov:2002:ALJJ:InsFluxon,Malomed:2004:ALJJ:Ic(Iinj)}, run around the LJJ periodically colliding with the fractional vortex. If the vortex's eigenfrequency is a multiple of the collision frequency, one should see  resonances on the IVC. The first study of a fluxon-semifluxon interaction in an annular LJJ was recently reported\cite{Goldobin:F-SF}, but the resonances, which appear due to the excitation of the eigenmodes, were not investigated in detail.

For future research, it is interesting to study an infinite array of fractional vortices (1D vortex crystal) which has only optical branches in the dispersion relation because of the vortex pinning. The energy bands of such an array can be tuned by changing $\kappa$ or bias current $\gamma$ \emph{during experiment}.

\begin{acknowledgments}
  We thank B. Malomed, A. Ustinov, A. Doelman, and G. Derks for fruitful discussions and suggestions. This work was supported by the Deutsche Forschungsgemeinschaft (GO-1106/1), by the ESF programs "Vortex" and "Pi-shift" and by the Royal Netherlands Academy of Arts and Sciences (KNAW). 
\end{acknowledgments}

\bibliography{MyJJ,LJJ,pi,SF,QuComp,SFS,software}

\end{document}